%% file: item_embedding.tex
\pdfoutput=1

\documentclass[sigconf]{acmart}

\usepackage{booktabs} 
\usepackage{color}
\usepackage{soul}
\usepackage[skip=5pt]{caption}

\setcopyright{rightsretained}

\acmConference[]{Under Review}{2019}{}
\copyrightyear{2019}

\begin{document}

\title{Personalized Ranking in eCommerce Search}

\author[G. Aslanyan, A. Mandal, P. Senthil Kumar, A. Jaiswal, M. R. Kannadasan]{Grigor Aslanyan, Aritra Mandal, Prathyusha Senthil Kumar, Amit Jaiswal, Manojkumar Rangasamy Kannadasan}
\affiliation{%
  \institution{eBay Inc.}
  \streetaddress{2025 Hamilton Avenue}
  \city{San Jose}
  \state{CA}
  \postcode{95125}
}
\email{gaslanyan,arimandal,prathykumar,amjaiswal,mkannadasan@ebay.com}

\input{Abstract}

\maketitle

\input{Introduction}
\input{Methodology}
\input{Results}
\input{Conclusion}

\bibliographystyle{ACM-Reference-Format}
\bibliography{bibliography}

\end{document}

%% file: Abstract.tex
\begin{abstract}
	We address the problem of personalization in the context of eCommerce search. Specifically, we develop personalization ranking features that use in-session context to augment a generic ranker optimized for conversion and relevance. We use a combination of latent features learned from item co-clicks in historic sessions and content-based features that use item title and price. Personalization in search has been discussed extensively in the existing literature. The novelty of our work is combining and comparing content-based and content-agnostic features and showing that they complement each other to result in a significant improvement of the ranker. Moreover, our technique does not require an explicit re-ranking step, does not rely on learning user profiles from long term search behavior, and does not involve complex modeling of query-item-user features. Our approach captures item co-click propensity using lightweight item embeddings. We experimentally show that our technique significantly outperforms a generic ranker in terms of Mean Reciprocal Rank (MRR). We also provide anecdotal evidence for the semantic similarity captured by the item embeddings on the eBay search engine.
\end{abstract}

%% file: Introduction.tex
\section{Introduction}\label{section-introduction}
In large eCommerce marketplaces that sell a wide variety of goods to very diverse buyer cohorts, there is no single best ordering of relevant results to search queries. As an example, for the query \textit{summer dress}, one user may prefer \textit{v-neck mini dresses} while another user may prefer \textit{floral maxi dresses}. As a result, a globally trained ranking model optimized for a generic notion of relevance or demand is incapable of capturing these specific user preferences. In order to provide a great shopping search experience, the search engine should be able to implicitly learn these preferences and leverage that understanding while surfacing search results.

Over the years, several techniques have been developed to incorporate personalization in web search \cite{dou2007large, bennett2012modeling, sontag2012probabilistic, wang2013personalized, grbovic2018real}. Most of these techniques use personalization signals to re-rank top N results from a generic search ranking model \cite{dou2007large, bennett2012modeling, sontag2012probabilistic} while a few of them directly incorporate those signals in the generic ranker \cite{grbovic2018real, wang2013personalized}. Another fundamental difference between published prior work in search personalization involves the distinction between content-based (query and document content) \cite{bennett2012modeling, sontag2012probabilistic} and user behavior based techniques \cite{cai2017behavior, yi2014beyond, shen2012personalized}, as well as hybrid techniques employing a combination of both. A number of these techniques rely on modeling a user profile or user specific features that rely on significant quantities of behavioral data for each user  \cite{teevan2005personalizing, sontag2012probabilistic, tan2006mining}. Alternatively, session based personalization techniques primarily rely on short term contextual features that are either user behavior based \cite{jiang2011context, shen2005context, xiang2010context} or content-based \cite{shen2005implicit, ustinovskiy2013personalization, mihalkova2009learning}.

More recently, following the success of deep neural networks for various NLP tasks, several approaches to personalization that rely on vector representations of queries, documents and/or users have been developed \cite{grbovic2018real, vu2017search, nguyen2018capsule, le2017personalization, ge2018personalizing, das2014commerce, palotti2016learning}. In particular, learning vector representations for documents (or items, in the case of eCommerce) and users using a sequence of user actions as context, in a way similar to embedding words using their context in sentences \cite{mikolov2013distributed, turian2010word, zhao2018learning, ai2017learning} has proven to be quite effective in several eCommerce applications including item recommendations \cite{wang2016learning, kenthapadi2017personalized}  and search \cite{grbovic2018real, rao2017personalization, le2017personalization}.

In this work we propose a search personalization technique that combines the effectiveness of several of the approaches described above using a simple formulation. Specifically, we leverage a combination of content-based and content-agnostic latent representation based features to personalize search results by directly using them as contextual features in the generic search ranker of eBay. For both types of features we use the most recent click context for the user in the current search session to derive these features. For the content-based features we utilize the title similarity and price similarity of the items to be ranked with those of the context items. These two content-based features are especially important for eCommerce purchase decisions. For the content-agnostic item embedding based features we learn low-dimensional item representations using their click context in historical search sessions. We then build personalization features by comparing these embeddings of the items to be ranked with the context items' representations similar to \cite{grbovic2018real}. We then augment our generic ranking model with these features and demonstrate the effectiveness of these personalization signals to improve MRR (Mean Reciprocal Rank) \cite{Craswell2009}. Our approach does not require an explicit re-ranking step, does not rely on user specific long term features, and does not involve complex modeling of query-item-user features to derive personalization features. We model item co-click propensity using simple and lightweight item embeddings and use a hybrid approach of mixing latent representations and explicit content-based features for personalization.

We also demonstrate qualitatively that the learned item embeddings capture item semantics very well - semantically similar items are close to each other in the embedding space even though none of these features were used for learning the embeddings. Therefore, the item embeddings can also be leveraged for use cases besides search ranking, such as recommendations, finding miscategorized items, detecting spam, and diversifying search result pages.

The main novelty of our work is combining and comparing content-based features with content-agnostic vector representation based features. We explicitly show that both types of features can significantly improve ranking for eCommerce search. We also show that vector representation based features complement content-based features and the combination of both gives the best results. To the best of our knowledge the application of this technique in the eCommerce domain is novel. We believe that this work can lead to more extensive future research, as discussed in Section \ref{section-summary}.

The rest of the paper is organized as follows. In Section \ref{section-method} we describe our methodology. In Section \ref{section-results} we present our results. We summarize our work and discuss future research in Section \ref{section-summary}.


%% file: Methodology.tex
\section{Methodology}\label{section-method}

\subsection{Item Embeddings}\label{subsection-embeddings}

We learn item embeddings from sequences of clicked items in search sessions. Each item on eBay has a unique ID and we use these IDs as our vocabulary. Our aim is to learn embeddings for each item ID. Importantly, the only signal that we use for training is user clicks. We do not include any information about the items themselves, such as title, image, price, or description.

We have collected data from one week of eBay search logs. For each search session we create a sequence of clicked item IDs which becomes a phrase. For example, if in a search session a user clicked on items $id_1$, $id_2$, and $id_3$ the phrase corresponding to that search event will be ``$id_1$ $id_2$ $id_3$''. We then keep all phrases with at least two words (i.e. 2 item IDs). Some of the item IDs appear only in one or just a few phrases making it difficult to train proper embeddings for them. For this reason we first filter out our vocabulary to keep only those item IDs that appear in at least $16$ phrases (the choice of this threshold is discussed in Section \ref{section-results}). In the phrases themselves we keep only those filtered item IDs in the vocabulary. With this setup we get a vocabulary size of about $8,000,000$ with nearly $85,000,000$ phrases in our data.

We use the publicly available \emph{word2vec} code for learning the embeddings using the skip-gram model \cite{mikolov2013distributed}. The input to the code is a text file which contains the phrases obtained as described above, one per line. We use hierarchical softmax for the objective function, window size of $5$, and $32$-dimensional embeddings. We also compare other choices for dimensionality.

\subsection{Ranking Features}\label{subsection-ranking-features}

A variety of personalization ranking features are developed based on the relationship between the items to be ranked in the current SRP (search results page) and up to 5 previously clicked items by the user in the same session. We develop two features that use item embeddings and two other features that use textual properties of items, namely price and title. The features that use embeddings are as follows:
\begin{itemize}
	\item \emph{\textbf{cos\_distance\_avg}}: Average cosine distance of the item to be ranked from the previously clicked items in the embedding space.
	\item \emph{\textbf{cos\_distance\_last}}: Cosine distance to the last clicked item with available embedding. If the last clicked item has an embedding then use that, otherwise the one before that, and so on, up to 5.
\end{itemize}

We also introduce the following features that use the price and title for comparing the current item to the previously clicked ones:
\begin{itemize}
	\item \emph{\textbf{price\_ratio\_mean}}: Ratio of the price of the current item to the average price of previously clicked items.
	\item \emph{\textbf{title\_jaccard\_sim}}: Jaccard similarity of the title of the current item to the last clicked item. Jaccard similarity is defined as the size of the intersection divided by the size of the union of the tokens of the two titles.
\end{itemize}

\subsection{Ranking Model}\label{subsection-ranking-model}

We train and evaluate ranking models with the features described in Section \ref{subsection-ranking-features} in addition to the features from the eBay generic ranking model that has no personalization features. Our goal is to evaluate if the proposed features are able to improve the ranking and to compare the personalization features that use embeddings to the hand-crafted features that use properties of the items like price and title.

We collect a sample of queries from one week of eBay search logs which is immediately after the week when the data to train the embeddings comes from. We use the LambdaMART algorithm \cite{from-ranknet-to-lambdarank-to-lambdamart-an-overview} for the ranking models with the same target as the eBay conversion ranking model. The baseline model uses the same ranking features as the production model (further details of the eBay ranking model are beyond the scope of this work). This means that the baseline model is similar to the eBay production ranking model, except that it uses less training data and different hyperparameters. We have used a fixed learning rate and have tuned the number of trees for each model using a separate validation dataset. We use MRR (mean reciprocal rank) \cite{Craswell2009} of sold items for evaluation.

In addition to the baseline model we train four new models which add the personalization features in Section \ref{subsection-ranking-features} to the features used in the baseline. Here is a description of the features used by all of the models:
\begin{itemize}
	\item \textbf{Baseline}: Same features as eBay production ranking model.
	\item \textbf{Distance\_Avg}: \textbf{Baseline} features + \emph{cos\_distance\_avg}.
	\item \textbf{Distance\_Last}: \textbf{Baseline} features + \emph{cos\_distance\_last}.
	\item \textbf{Price\_Title}: \textbf{Baseline} features + \emph{price\_ratio\_mean} + \\   \emph{title\_jaccard\_sim}.
	\item \textbf{All}: \textbf{Baseline} features + all of the proposed personalization features.
\end{itemize}

The embeddings that we learn do not cover the entire inventory. Since our training and test datasets are collected using a fair sample of queries we do not get $100\%$ coverage for embeddings for the items to be ranked, as well as previously clicked items by the user. To get a fair evaluation of the proposed ranking features that use embeddings and a fair comparison with the hand crafted features we filter out the items that do not have an embedding, as well as queries where none of the last 5 clicked items has an embedding. We also filter out items for which the last clicked item with an available embedding is the same as the current item. This is done to ensure that we evaluate the value of embeddings for ranking rather than a feature that simply identifies if the current item was among the last 5 clicked items. The filtering is done for both the training and test datasets. After that filtering we keep only queries that have at least one positive item and a sufficient total number of items left in the data (at least 3 items total per query for training and at least 20 per query for test). We will refer to this dataset as the \textbf{high coverage} dataset.

Separately, we have also trained and evaluated ranking models on the full datasets. Even though we do not get full coverage for embeddings we still want to evaluate the usefulness of the proposed ranking features on the full data.

%% file: Results.tex
\section{Results}\label{section-results}

\begin{figure}
	\includegraphics[height=2.3cm, width=8.5cm]{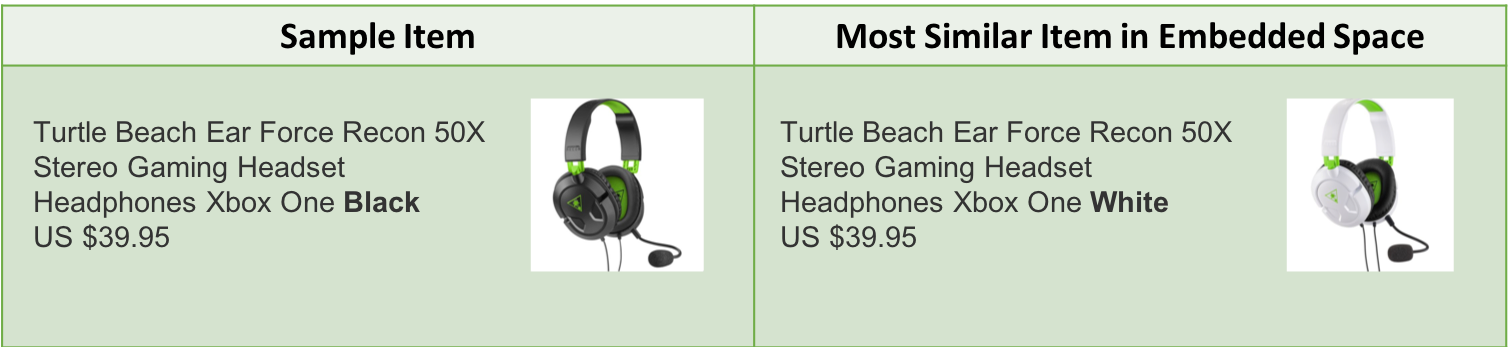}
	\setlength{\belowcaptionskip}{-12pt}
	\caption{A sample item from eBay on the left, compared to its most similar item in the embedding space on the right. The items were listed by the same seller and only differ in color. Note that when training the embeddings no properties of the items were used, like title, price, or seller.}\label{fig-item-similarity}
\end{figure}

We first perform a qualitative evaluation on the item embeddings to verify if similar items in the embedding space are semantically and visually similar. Fig.~\ref{fig-item-similarity} shows an example based on embeddings of dimension 32, trained using a window size 5. Given that we have not used any item features for training the embeddings (title, price, images, seller) this is a very encouraging result. We have noticed similarly good results for most of the items we have examined.

\begin{figure}
	\includegraphics[height=5cm, width=8cm]{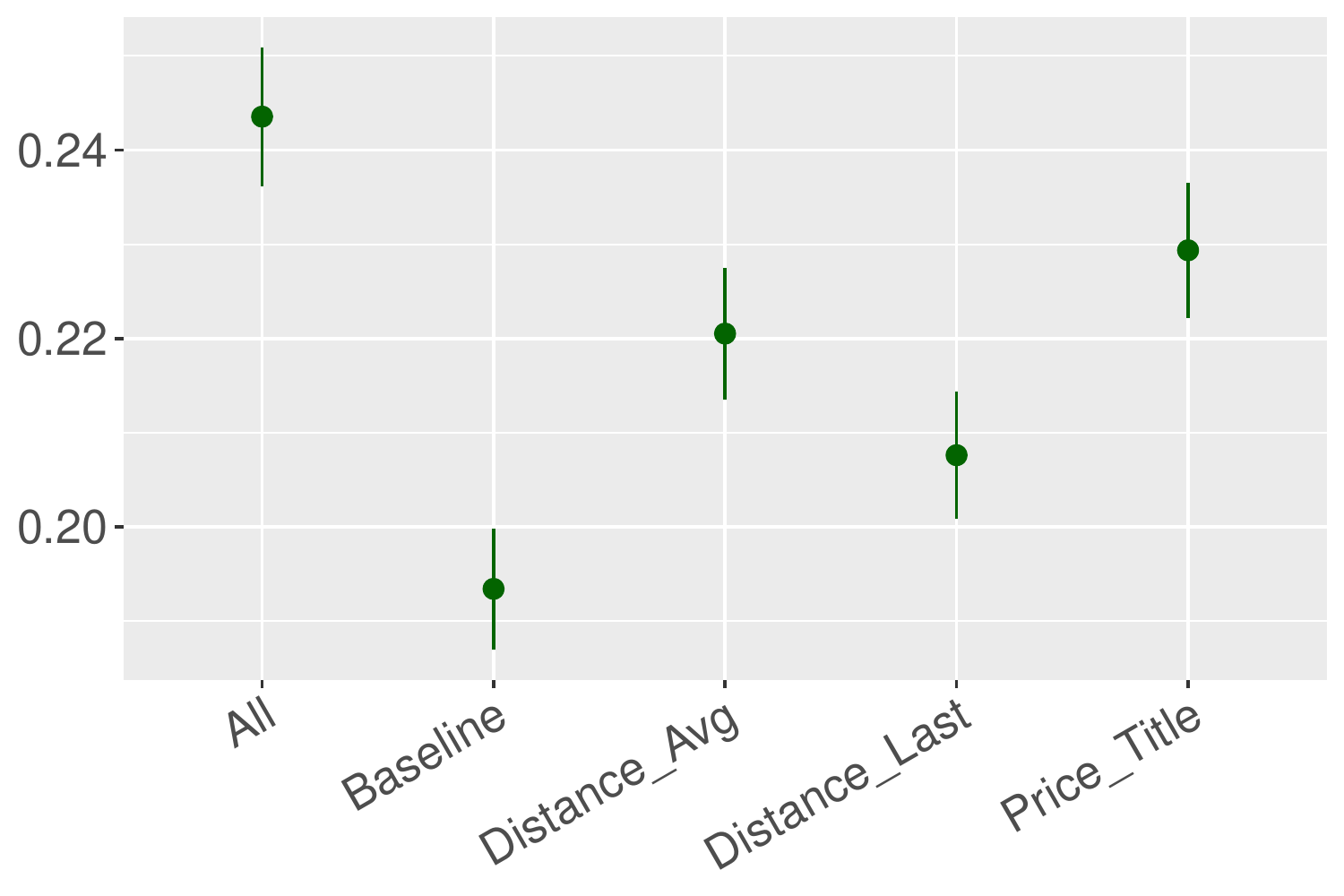}
	\setlength{\belowcaptionskip}{-12pt}
	\caption{MRR for the learning to rank models on the high coverage dataset.}\label{fig-mrr-sale}
\end{figure}

MRR for the ranking models is plotted in Fig.~\ref{fig-mrr-sale}. We have performed a bootstrap over the queries in the test set and plotted the median together with $95\%$ confidence intervals. This allows us to verify that improvements are statistically significant. These models were trained and tested on the \textbf{high coverage} dataset (it includes full coverage for the proposed ranking features and excludes items that coincide with one of the last clicked ones). This gives us a fair comparison and evaluation for the new features. The \emph{cos\_distance\_last} feature gives an improvement in MRR of $8\%$, while \emph{cos\_distance\_avg} results in a $15\%$ improvement. The hand crafted features by themselves result in a slightly higher improvement - $19\%$. By combining all of the features we get an even higher improvement in MRR - $26\%$. This means that the embedding based features and hand crafted features complement each other. Even though hand crafted features perform slightly better than embedding based features, the embeddings are still able to add more value.

\begin{figure}
	\includegraphics[height=5cm, width=8cm]{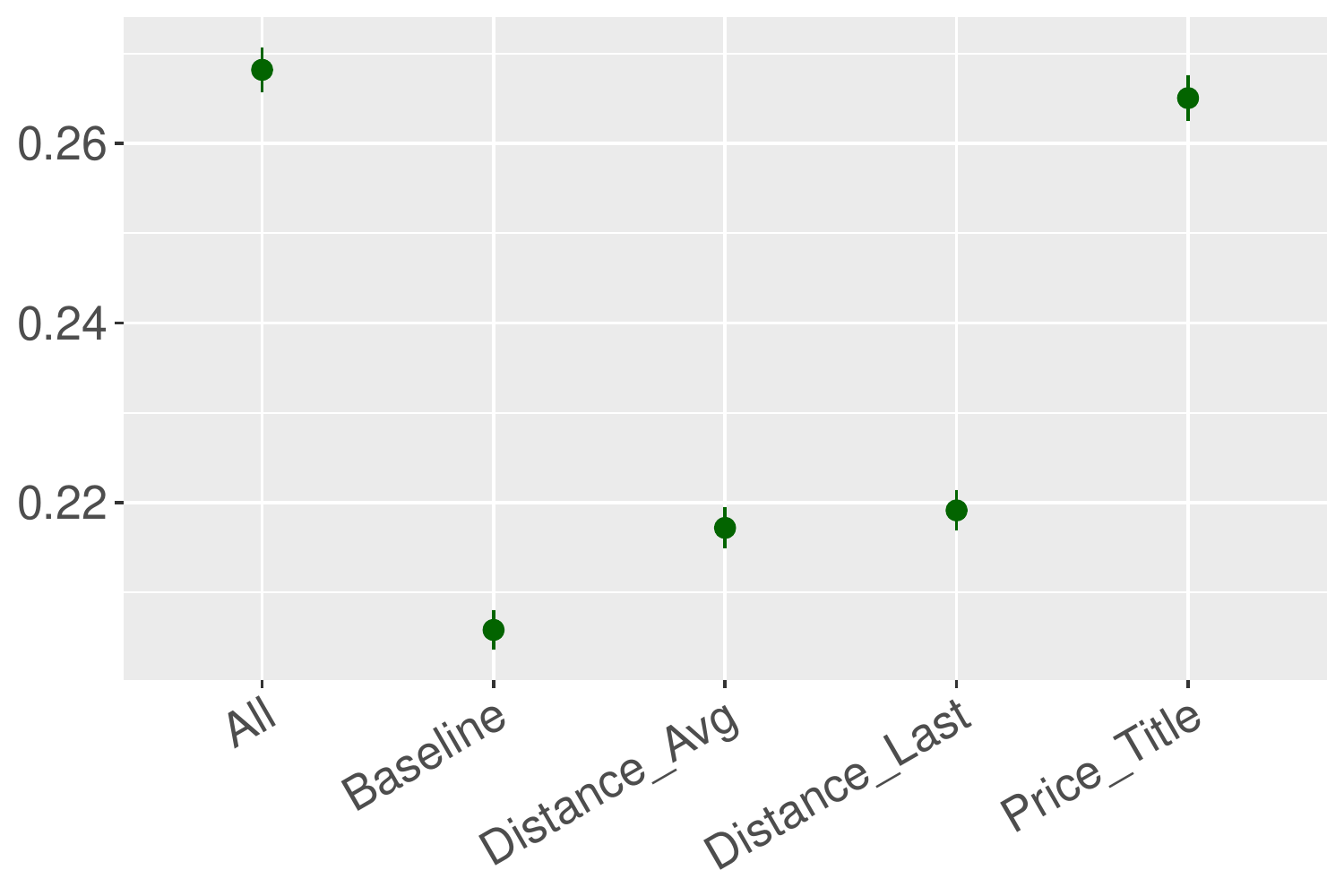}
	\setlength{\belowcaptionskip}{-12pt}
	\caption{MRR for the learning to rank models on the full dataset. The coverage for embedding based features is $8\%$.}\label{fig-mrr-sale-full}
\end{figure}

We have also explored these ranking features on the full training and test datasets, which contains a fair sample of search queries. For this data we get about $14\%$ coverage for embeddings. However, the coverage for the ranking features \emph{cos\_distance\_last} and \emph{cos\_distance\_avg} is even lower - $8\%$. This is because these features also require embeddings coverage for at least one of the last 5 clicked items by the user. The coverage for the hand crafted features \emph{price\_ratio\_mean} and \emph{title\_jaccard\_sim} is much higher - around $80\%$. MRR for the models with the full data is plotted in Fig.~\ref{fig-mrr-sale-full}. As expected, we get a much larger improvement by using the hand crafted features than the embedding based features - the hand crafted features give an improvement of $29\%$ while the embedding based features result in a $6\%$ improvement\footnote{As opposed to the \textbf{high coverage} dataset we have not filtered out items from the full dataset that coincide with one of the last clicked items. This is the reason why we get a larger overall improvement for the full dataset. We have simply excluded these items from the high coverage dataset since that dataset is used to evaluate and compare the features themselves rather than their ability to identify a coinciding item among the recent clicks.}. This is because of the large difference in coverage. Nevertheless, even with under $10\%$ coverage the embedding based features are able to bring an improvement to the model. Also, when combining all of the features we get an extra $1\%$ improvement compared with the model that uses hand crafted features only.

We have explored different thresholds for the number of clicks that an item must have received to be included into our embeddings vocabulary (default value $16$). However, by relaxing this threshold we have not seen significantly larger coverage of embeddings in the training data and no significant difference in MRR.

\begin{table}
	\begin{center}
		\caption{MRR improvement compared to the baseline model as a function of the dimensionality of the embedding space for the high coverage data. The ``32 dim.'' column corresponds to the results in Fig.~\ref{fig-mrr-sale}.}
		\label{dim-table}
		\begin{tabular}{|c|c|c|c|c|}
			\hline
			Model & 32 dim. & 16 dim. & 8 dim. & 4 dim.  \\
			\hline
			Distance\_Avg & $15\%$ & $10\%$ & $7\%$ & $4\%$  \\
			\hline
			Distance\_Last & $8\%$ & $5\%$ & $4\%$ & $2\%$  \\
			\hline
			All & $26\%$ & $24\%$ & $19\%$ & $19\%$  \\
			\hline
		\end{tabular}
	\end{center}
	\vspace{-6mm}
\end{table}

We have also explored how the dimensionality of the embeddings space affects the ranking features and MRR. The improvements in MRR for different dimensions of the embedding space are shown in Table \ref{dim-table}. As expected, MRR improvements decrease with decreasing dimensionality. At $8$ dimensions and below the added value from embeddings in addition to the hand crafted features becomes insignificant.

%% file: Conclusion.tex
\section{Summary and Future Work}\label{section-summary}

In this work we have presented our approach to incorporating personalization in eCommerce search by developing a combination of content-based features and content-agnostic latent vector representation based features. These contextualization features are directly included in the generic search ranker. We have applied our technique on a large commercial eCommerce search engine (eBay) and have obtained significant improvements in offline ranking metrics. Specifically, when restricting the dataset to have high coverage for the proposed features we have seen an improvement of $15\%$ in MRR by using only embedding based features, and $26\%$ when combining them with content based features. We have also seen significant improvements in MRR on the full dataset. However, on the full dataset the content based features have a larger impact due to much larger coverage. Nevertheless, the embedding based features still add significant value even though the coverage for them is under $10\%$ in this dataset. An important future area of research is to improve this coverage, which includes training embeddings on longer click sequences (beyond a search session). It would also be useful to go beyond the last 5 clicks for personalization ranking features and explore longer term user activity, as well as user activity by category.

We anecdotally demonstrate the effectiveness of our item embeddings in capturing semantic properties of the items, even though no properties of the items were used in learning the embeddings. This indicates that the embeddings could be valuable in a number of other user cases besides search ranking, such as recommendations, finding miscategorized items, detecting spam, and diversifying the search result pages. We plan to explore these areas in future work.